# Preemptive Two-stage Goal-Programming Formulation of a "Strict" Version of the Unbounded Knapsack Problem with Bounded Weights

Mike Beyer, Steven Mills

**Abstract**

The unbounded knapsack problem with bounded weights is a variant of the well-studied variant of the traditional binary knapsack problem; key changes being the relaxation of the binary constraint and allowing the unit weights of each item to fall within a range. In this paper, we formulate a variant of this problem, which we call the "strict" unbounded knapsack problem with bounded weights, by replacing the inequality constraint on the total weight with an equality. We show that this problem can be decomposed into a two-stage, pre-emptive goal programming problem, with the first stage being a 2-dimensional knapsack problem and the second being either a linear feasibility program (per canonical formulation) or simply a linearly-constrained program in the general case. This reformulation is shown to be equivalent to the original formulation but allows the use of well-studied, efficient algorithms for multidimensional knapsack problems. In addition, it separates the modeling effort around *what* to put in the knapsack from considerations around what *unit weight* one should assign to each item type, providing substantially more flexibility to the modeler without adding complexity to the choice of knapsack configuration. Finally, we show that for the feasibility version of the second stage, one can immediately get a feasible solution to the first stage solution.

**Keywords** knapsack goal programming preemptive multidimensional multiconstraint unbounded

## Introduction

Knapsack problems are well-known combinatorial optimization problems that arise in many real-world settings, including capital budgeting, location, and allocation [1]. Given a set of *n* items, each with a known weight $w_i$ and value $v_i$, the goal of the knapsack problem is to find the set of items with maximal total value and a total weight that does not exceed a specified limit *W*. At most one of any item can be selected. The unbounded knapsack problem is a variant that relaxes this final limitation, instead allowing an unbounded, non-negative number of each item type to be selected. The unbounded knapsack is formally defined as:

$$Maximize \sum_{i=1}^{n} v_i x_i$$

$$s.t. \sum_{i=1}^{n} w_i x_i \leq W$$

$$x_i \in \mathbb{N}_0 \quad \forall \quad i = 1, \dots, n$$

In this case, the $x_i$ no longer represent individual items, but item *types*, each with a fixed *unit weight* $w_i$. However, there are practical applications where the unit weights are not fixed, but can instead be considered as part of the decision process. In this paper, we introduce a new variant of the unbounded knapsack problem, motivated by use cases where weights are variable but must fall within pre-specified bounds. Furthermore, rather than the total weight falling below a specific limit *W*, we enforce strict equality of the total weight. We call this variant the *strict* *unbounded knapsack with bounded weights (UKBW-s)*, where solutions are constrained to the hypersurface of constant total knapsack weight.

The remainder of this paper is organized as follows: Section 1 formally defines the UKBW-s as a mathematical programming problem. Section 2 compares it to similar variants of the knapsack problem. Section 3 details a two-stage preemptive goal programming formulation of UKBW-s.

## 1. Formal problem definition

Given a set of *n* item types, each with a known value $v_i$ and a variable weight for each item type $w_i$ that must fall within the interval ($w_i^{min}$, $w_i^{max}$), select the number of each item type and the corresponding weights to maximize total value while exactly matching a specified total weight *W*. The strict unbounded knapsack with bounded weights (UKBW-s) is canonically defined as:

$$Maximize \sum_{i=1}^{n} v_i x_i$$

$$s.t. \sum_{i=1}^{n} w_i x_i = W$$

$$Where, for\ i = 1, \ldots, n:$$
$$w_i^{min} \leq w_i \leq w_i^{max}$$
$$x_i \in \mathbb{N}_0$$
$$v_i, w_i^{min} > 0$$

## 2. Related work

UKBW-s has four defining characteristics.
1. The number of each item type is unbounded,
2. Unit-weights are decision variables,
3. Unit-weights must fall within pre-specified bounds, and
4. Strict equality on total weight constraint.

While several related problems have been studied in the literature, none capture this unique combination of attributes. In fact, UKBW-s combines the main attributes of two related variants of the classic knapsack problem, namely the Change-making Problem (CMP) and the Knapsack under Linear Constraints (KLC). We can observe this relationship from the table of attributes below. In the rest of this section, we will mainly focus on comparing UKBW-I with KLC and CMP. We will also briefly discuss the relationship of UKBW-s with a transitional model called the Robust Knapsack Problem (RKP).

Table 1 summarizes the combination of problem attributes that are unique to UKBW-I and compares them across these related works.

| Problem Attribute | CMP | RKP | KLC | UKBW-I |
|---|---|---|---|---|
| Unbounded knapsack | ✓ | | | ✓ |
| Strict equality on total weight | ✓ | | | ✓ |
| Weights as decision variables | | | ✓ | ✓ |
| Weights bounded on interval | | ✓ | ✓ | ✓ |

Table 1. Comparison of key problem attributes across related works

Martello and Toth introduce the Change-making Problem (CMP), which aims to minimize the number of coins required to reach a defined monetary value. The problem is unbounded in that it allows an unlimited number of each coin type. Importantly, strict equality of total weight (the total monetary value of change required) is enforced. The weights are fixed, however, in that each coin type has a predefined monetary value.

The Robust Knapsack Problem (RKP) addresses uncertain weights. It is a variant on the original 0-1 knapsack problem where values are fixed, but weights are uncertain and bounded on the interval $[w_i - w_i^l, w_i + w_i^u]$. The objective of RKP is to find a solution which maximizes the total value and is robust to the imprecise weights. An important distinction relative to the UKBW-s is that, in the RKP, the weights are not part of the decision process. While the weights may vary due to uncertainty around their specific value, they remain parameters that act as coefficients in the problem formulation.

The Knapsack under Linear Constraints (KLC) is a generalization of the classical 0-1 knapsack problem. Unlike the RKP, KLC includes weights $w_i$ as explicit decision variables and introduces a set of linear constraints on $w_i$. UKBW-s is similar in this way, but UKBW-s differs in that it is based on the unbounded knapsack, relaxing the 0-1 requirement on the number of each item type. Furthermore, the UKBW-s is a special case where the weights are bounded within an interval and strict equality is enforced on total weight.

## 3. Two-stage, preemptive goal programming formulation

The canonical formulation of UKBW-s is a single-stage mixed-integer program with discrete item counts, continuous weights, and a bilinear equality constraint. However, the reader may have noticed that the variable weights do not appear in the objective function; instead, they exert their effect indirectly by constraining the allowable knapsack configuration $x$. This implies that the solution may be underdetermined: for the item types included in a given optimal knapsack configuration $x^*$, there may be multiple (possibly uncountably infinite) values for one or more pairs of values in $w^*$. A sufficient condition for this to occur is when one item type included in $x^*$ has upward slack in its weight constraints and another item type included in $x^*$ has downward slack in its weight constraints.

**Definition 3.1** *Given a knapsack configuration x, let $J(x) := \{j : x_j > 0\}$ be the set of item types included in the knapsack.*

**Theorem 3.2** *Given an optimal knapsack configuration $x^*$, if $\exists i, j \in J(x^*): w_i^{max} - w_i^* > 0$ and $w_j^* - w_j^{min} > 0$, then $(x^*, w^*)$ is not unique.*

*Proof* Given an optimal knapsack configuration $x^*$, we know that $\exists w^*: \sum_{d \in J(x^*)} x_d^* w_d^* = W$ by the definition of $J(x^*)$. Assume $\exists i, j \in J(x^*): w_i^{max} - w_i^* > 0 \text{ and } w_j^* - w_j^{min} > 0$.

Let
$D = x_i^* w_i^* + x_j^* w_j^*$ be the total weight of items of type $i$ or $j$
$\gamma := \frac{x_j^*}{x_i^*}$ be the $j$-$i$ count ratio
$\Delta := \min\{w_i^{max} - w_i^*, \gamma(w_j^* - w_j^{min})\}$ be the maximum allowable *unit* weight increase for items of type $i$.
$R = W - D$ be the remaining knapsack weight after removing items of type $i$ or $j$

We then have $R + D = R + w_i^* x_i^* + w_j^* x_j^* = W \Rightarrow R + (w_i^* + \delta)x_i^* + \left(w_j^* - \frac{\delta}{\gamma}\right)x_j^* = R + w_i^* x_i^* + \delta x_i^* + w_j^* x_j^* - \frac{\delta}{\gamma} x_j^* = R + w_i^* x_i^* + \delta x_i^* + w_j^* x_j^* - \delta x_i^* = W \quad \forall \delta \in [0, \Delta]$

That is, $(x^*, \tilde{w}(\delta))$ are also optimal solutions for all $\delta \in [0, \Delta]$, where $\tilde{w}(\delta) = w^* + \delta x_i - \frac{\delta}{\gamma} e_j$. Since $Card([0, \Delta]) = \mathfrak{c}$, there are an uncountably infinite number of optimal solutions. ∎

The degeneracy in the weights will result in arbitrary weight vectors being chosen from among the optimal solutions. One remedy would be to add additional weight terms to the objective function to enforce uniqueness; however, this implies that item values and weight "penalties" are fungible, which not only can be conceptually unsound, but fundamentally changes the problem. A second approach, and the one advocated here, is to use a two-stage, preemptive goal programming formulation: a configuration stage (Stage 1) and a weight-selection stage (Stage 2), where the optimal configuration is preemptive over weight selection. As will be shown in the following sections, this re-formation is equivalent to the original problem but shifts the weight degeneracy of (3.2) out of the main problem while also allowing additional modeling flexibility in selecting the weights to assign the optimal item types. A key computational benefit of this reformulation is it allows Stage 1 to be expressed as a multidimensional knapsack problem, a well-studied NP-hard problem with a number of effective exact and approximate solution algorithms (Puchinger, Raidl, Pferschy, 2010; Varnamkhasti 2012). The main conceptual benefit is the separation of concerns between configuration selection (Stage 1) and weighting preferences (Stage 2).

### *Stage 1: Knapsack Configuration*

Formulating the configuration stage problem requires removing the weight variables from the canonical formulation. This can be done by replacing the equality constraint on the total weight by two inequality constraints, resulting in a multidimensional unbounded knapsack problem.

$$\text{Maximize} \sum_{i=1}^{n} v_i x_i$$

$$s.t.$$

$$\sum_{i=1}^{n} w_i^{max} x_i \geq W$$

$$\sum_{i=1}^{n} w_i^{min} x_i \leq W$$

$$x_i \in \mathbb{N}_0 \; i = 1, \ldots, n$$

The two inequality constraints can be interpreted as saying that the range of possible total weights for the knapsack (given configuration $x$) must include W. Informally, this relies upon the fact that each of the weight variables are (at least nominally) continuous over some interval of $\mathbb{R}_+$. This will be expanded upon below.

**Definition 3.3** Let $Q := [w_1^{min}, w_1^{max}] \times \ldots \times [w_n^{min}, w_n^{max}]$ be the set of allowed values of $w$.

**Definition 3.4** Let $K(x, w) : \mathbb{N}_0^n \times \mathbb{R}_+^n \to \mathbb{R}_+ \; ; (x, w) \mapsto \langle x, w \rangle$, be the total weight of a knapsack with configuration $x$ and weights $w$.

**Definition 3.5** The weight span $S_Q(x)$ of a given knapsack configuration $x$, is the image of $K(x, w)$ over the restricted domain $x \times Q$.

**Lemma 3.6** $S_Q(x)$ is a compact subset of $\mathbb{R}_+$.

*Proof* $Q$ is (trivially) a closed and bounded subset of $\mathbb{R}_+^n$; therefore, it is compact. In addition, $K(x, w)$ (with $x$ fixed) is a linear (and hence continuous) function from $\mathbb{R}_+^n$ to $\mathbb{R}_+$. Compact sets are preserved under continuous transformations; therefore, $S_Q(x)$ is also a compact set. ∎

**Definition 3.7** Let $\overline{w} := (w_1^{max}, \ldots, w_n^{max})$ and $\underline{w} := (w_1^{min}, \ldots, w_n^{min})$ be the extreme weight vectors.

**Lemma 3.8** $\min S_Q(x) = K(x, \underline{w})$ and $\max S_Q(x) = K(x, \overline{w})$

*Proof* By Definition 3.5, $K(x, \underline{w})$ and $K(x, \overline{w})$ are members of $S_Q(x)$; we will demonstrate that they are also the $\inf S_Q(x)$ and $\sup S_Q(x)$, respectively. For $\min S_Q(x)$: $\inf S_Q(x) < K(x, \underline{w}) \Rightarrow \exists i : w_i < w_i^{min}$, which is contradiction. In addition, $\inf S_Q(x) > K(x, \underline{w})$ is trivially false; therefore, $\inf S_Q = K(x, \underline{w}) \in S_Q(x) \Rightarrow \min S_{Q(x)} = K(x, \underline{w})$. A similar argument holds for showing $\max S_{Q(x)} = K(x, \overline{w})$ ∎

**Corollary 3.9** $S_Q(x) = [K(x, \underline{w}), K(x, \overline{w})]$
*Proof* This follows directly from (3.6) and (3.8).

**Theorem 3.10** $K(x,\underline{w}) \leq W \leq K(x,\overline{w}) \implies \exists w' \in Q : K(w', x) = W$

*Proof* This follows directly from (3.9) and (3.5)

(3.10) ensures that a feasible solution $x$ in Stage 1 will have at least one corresponding feasible solution $(w, x)$ in the original problem. The reverse is also true – let $(w, x)$ be a feasible solution to the *canonical* formulation of the problem. If we take the scalar product of the unit weight constraint of the canonical problem (in vector form), $0 < w^{min} \leq w \leq w^{max}$, with $x$ we see that that $x$ also satisfies the constraints of Stage 1:

$$(x > 0) \text{ and } (0 < w^{min} \leq w \leq w^{max}) \implies w^{min} \cdot x \leq w \cdot x \leq w^{max} \cdot x$$
$$\implies w^{min} \cdot x \leq W \leq w^{max} \cdot x \quad \blacksquare$$

Therefore, every feasible configuration in the canonical problem is also a feasible configuration in Stage 1 and every feasible configuration in Stage 1 can be paired with at least one feasible weight vector in the canonical problem to produce a feasible pair of weight-configuration vectors. This implies that the set of feasible configurations are identical for both the original problem and Stage 1. Since the objective functions are identical, the set of optimal configurations will be the same as well.

With the optimal configuration selected in Stage 1, we can move on to Stage 2, which is focused on identifying unit weights for the item types included in the knapsack from Stage 1.

### Stage 2: Weight vector selection

The general formulation of Stage 2 is a mathematical program with a single linear equality constraint over $Q$ evaluated using the Stage 1 solution $x^*$.

$$\text{Maximize } f(w; x^*)$$
$$s.t.$$
$$\sum_{i=1}^{n} w_i x_i^* = W$$
$$w_i \in Q_i$$

The existence of a unique optimal solution to this problem depends on the form of $f(w; x^*)$. The original form of the UKBW-s problem is equivalent to solving the two-stage formulation using $f(w; x^*) \coloneqq c$, for any $c \in \mathbb{R}$ in Stage 2, reducing it to a feasibility problem. As with the original formulation, a unique solution to the weights is not guaranteed in this case; however, there is an easy "default" solution that is guaranteed to exist (as we proved in (3.10)). Therefore, even in situations where one does not require uniqueness, the two-stage formulation has the benefit of allowing an efficient optimal solution to Stage 1 and a simple formula for a feasible solution to Stage 2.

*Guaranteed feasible point*

As indicated above, in the case where $f(w; x^*)$ is a constant we can skip the Stage 2 mathematical program (if we are not particular about the values of the components of $w$). If specific weights are needed, then the following solution is guaranteed to be feasible if $x^*$ exists.

**Definition 3.11** $\sigma := \frac{W - \langle \underline{w}, x^* \rangle}{\langle \overline{w}, x^* \rangle - \langle \underline{w}, x^* \rangle}$ is the relative position of the target total weight value $W$ along the weight span for solution $x^*$.

**Theorem 3.12** $w_i = w_i^{min} + \sigma(w_i^{max} - w_i^{min})$ is a feasible solution to the Stage 2 problem.

*Proof* As we assumed that the stage 1 solution exists – that is, $K(x^*, \underline{w}) \leq W \leq K(x^*, \overline{w})$ or equivalently $\langle \underline{w}, x^* \rangle \leq W \leq \langle \overline{w}, x^* \rangle$, we have from Theorem 3.10 that there exists $w' \in Q$ such that $\langle w', x^* \rangle = W$. Together, this means $\langle \underline{w}, x^* \rangle \leq \langle w', x^* \rangle \leq \langle \overline{w}, x^* \rangle$. Using Corollary 3.9 we have $\langle w', x^* \rangle \in S_Q(x^*)$. Since $S_Q(x^*)$ is a compact subset of $\mathbb{R}_+$ as a result of Lemma 3.6, we can write $\langle w', x^* \rangle = \sigma \langle \overline{w}, x^* \rangle + (1 - \sigma)\langle \underline{w}, x^* \rangle$ as a convex combination of $\langle \overline{w}, x^* \rangle$ and $\langle \underline{w}, x^* \rangle$ for some $\sigma \in [0,1]$, which rearranges to $\langle w', x^* \rangle = \langle \underline{w}, x^* \rangle + \sigma(\langle \overline{w}, x^* \rangle - \langle \underline{w}, x^* \rangle)$. Solving for $\sigma$ yields

$$\sigma = \frac{\langle w', x^* \rangle - \langle \underline{w}, x^* \rangle}{\langle \overline{w}, x^* \rangle - \langle \underline{w}, x^* \rangle} = \frac{W - \langle \underline{w}, x^* \rangle}{\langle \overline{w}, x^* \rangle - \langle \underline{w}, x^* \rangle}$$

Substituting this back we get

$$W = \langle w', x^* \rangle = \langle \underline{w}, x^* \rangle + \sigma(\langle \overline{w}, x^* \rangle - \langle \underline{w}, x^* \rangle) = \langle \underline{w} + \sigma(\overline{w} - \underline{w}), x^* \rangle$$

This means that $w^* = \underline{w} + \sigma(\overline{w} - \underline{w})$ where $\sigma = \frac{W - \langle \underline{w}, x^* \rangle}{\langle \overline{w}, x^* \rangle - \langle \underline{w}, x^* \rangle}$ is a feasible solution.[1]

**Acknowledgements**

We would like to thank Mr. Tianyang (Tim) Yang and Dr. Arjumand Masood for their contributions to earlier versions of this paper and the associated computational studies. We would also like to thank Mr. Liam Connell, Mr. Tobias Holler, and Dr. Kelvin Hsu for providing helpful review and feedback on the proofs.

---

[1] We thank Dr. Kelvin Hsu for suggesting this version of the proof.